\documentclass[prl,twocolumn,nofootinbib,superscriptaddress]{revtex4-1}

\usepackage{amsfonts}
\usepackage{amsmath}
\usepackage{amssymb}
\usepackage{amsthm}
\usepackage{bm}
\usepackage{dcolumn}
\usepackage{epsfig}
\usepackage{graphicx}
\usepackage{graphics}
\usepackage[latin1]{inputenc}
\usepackage{latexsym}
\usepackage{rotating}

\begin{document}

\title{Learning about black hole binaries from their ringdown spectra}

\author{Scott A.\ Hughes} \affiliation{Department of Physics and MIT Kavli Institute, Massachusetts Institute of Technology, Cambridge, MA 02139}

\author{Anuj Apte} \affiliation{Department of Physics and MIT Kavli Institute, Massachusetts Institute of Technology, Cambridge, MA 02139}

\author{Gaurav Khanna} \affiliation{Department of Physics, University of Massachusetts, Dartmouth, MA 02747}

\author{Halston Lim} \affiliation{Department of Physics and MIT Kavli Institute, Massachusetts Institute of Technology, Cambridge, MA 02139}

\begin{abstract}
The coalescence of two black holes generates gravitational waves that carry detailed information about the properties of those black holes and their binary configuration.  The final coalescence cycles are in the form of a {\it ringdown}: a superposition of quasi-normal modes of the merged remnant black hole.  Each mode has an oscillation frequency and decay time that in general relativity is determined by the remnant's mass and spin.  Measuring the frequency and decay time of multiple modes makes it possible to measure the remnant's mass and spin, and to test the waves against the predictions of gravity theories.  In this {\it Letter}, we show that the relative amplitudes of these modes encodes information about a binary's {\it geometry}.  Focusing on the large mass-ratio limit, which provides a simple-to-use tool for effectively exploring parameter space, we demonstrate how a binary's geometry is encoded in the relative amplitudes of these modes, and how to parameterize the modes in this limit.  Although more work is needed to assess how well this carries over to less extreme mass ratios, our results indicate that measuring multiple ringdown modes from coalescence may aid in measuring important source properties, such as the misalignment of its members' spins and orbit.
\end{abstract}

\maketitle

\noindent {\it Introduction.}  Binary black hole systems have, so far, proven to be the most commonly measured gravitational-wave (GW) sources \cite{gwtc1}.  A combination of analytic and numerical techniques makes it possible to model the waveform produced by two black holes as they inspiral and merge into a single hole.  These models are invaluable to the analysis of their GWs, providing a parameterized framework that facilitates finding signals in noisy detector data and makes it possible to characterize the system which produced the waves.

The final GW cycles from merging black holes are a superposition of {\it ringdown} modes.  Within general relativity, mode properties are set by the ringing hole: each mode has a frequency and decay time determined by the hole's mass and spin, and each radiates into an angular pattern that depends on the hole's spin, the mode's frequency, and its angular indices $(\ell, m)$ {\cite{qnmreview}}.  By measuring multiple modes and assuming general relativity, ringdown waves offer a means of measuring the final black hole mass and spin {\cite{bhspectro1}}.  By relaxing the assumption of general relativity, one can use these modes {\cite{bhspectro2}} as a tool for checking that the waves' properties are consistent with the Kerr solution for rotating black holes {\cite{kerr63}}.

\begin{figure*}
\includegraphics[width = 0.2798\textwidth]{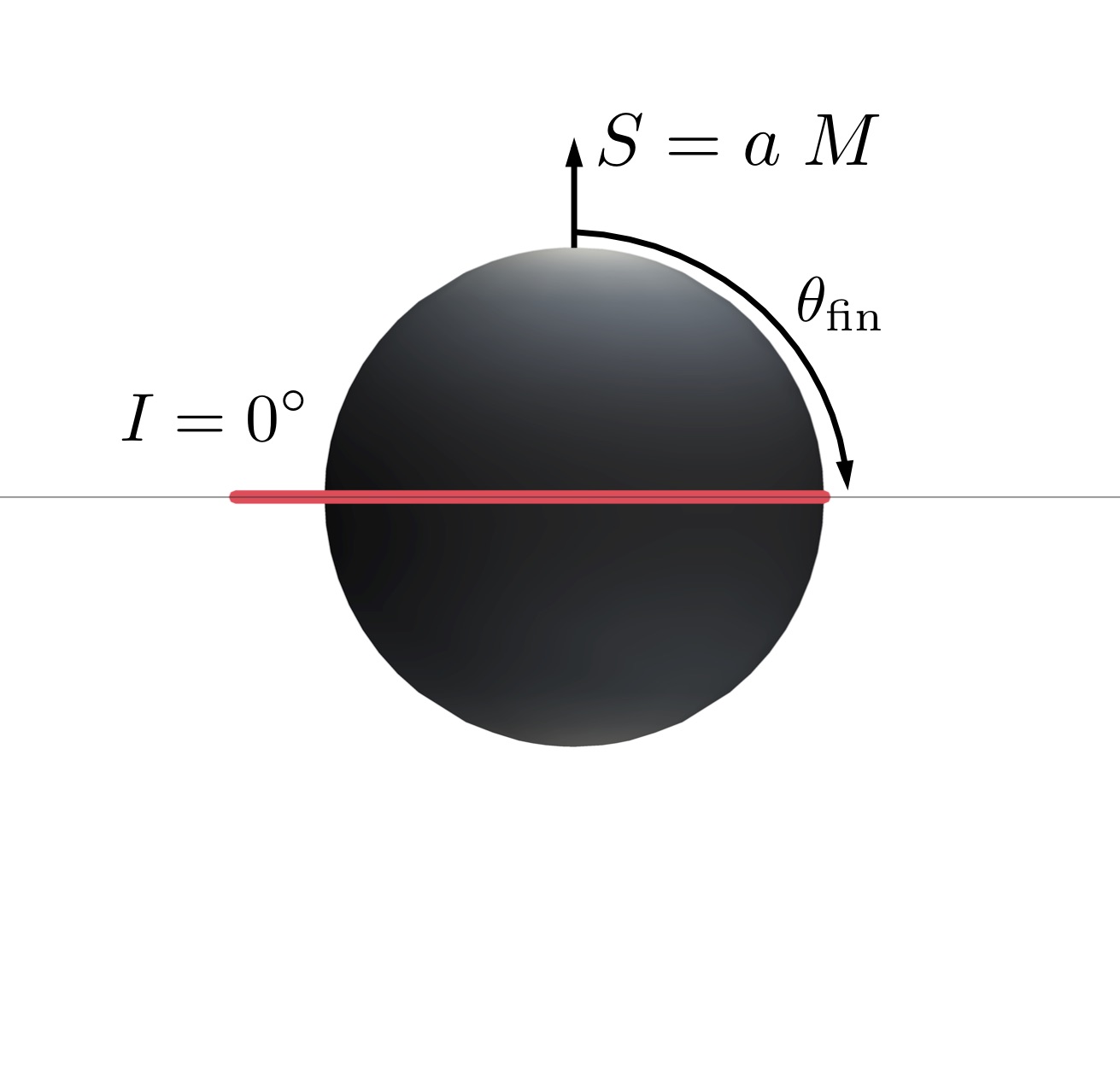}
\includegraphics[width = 0.2978\textwidth]{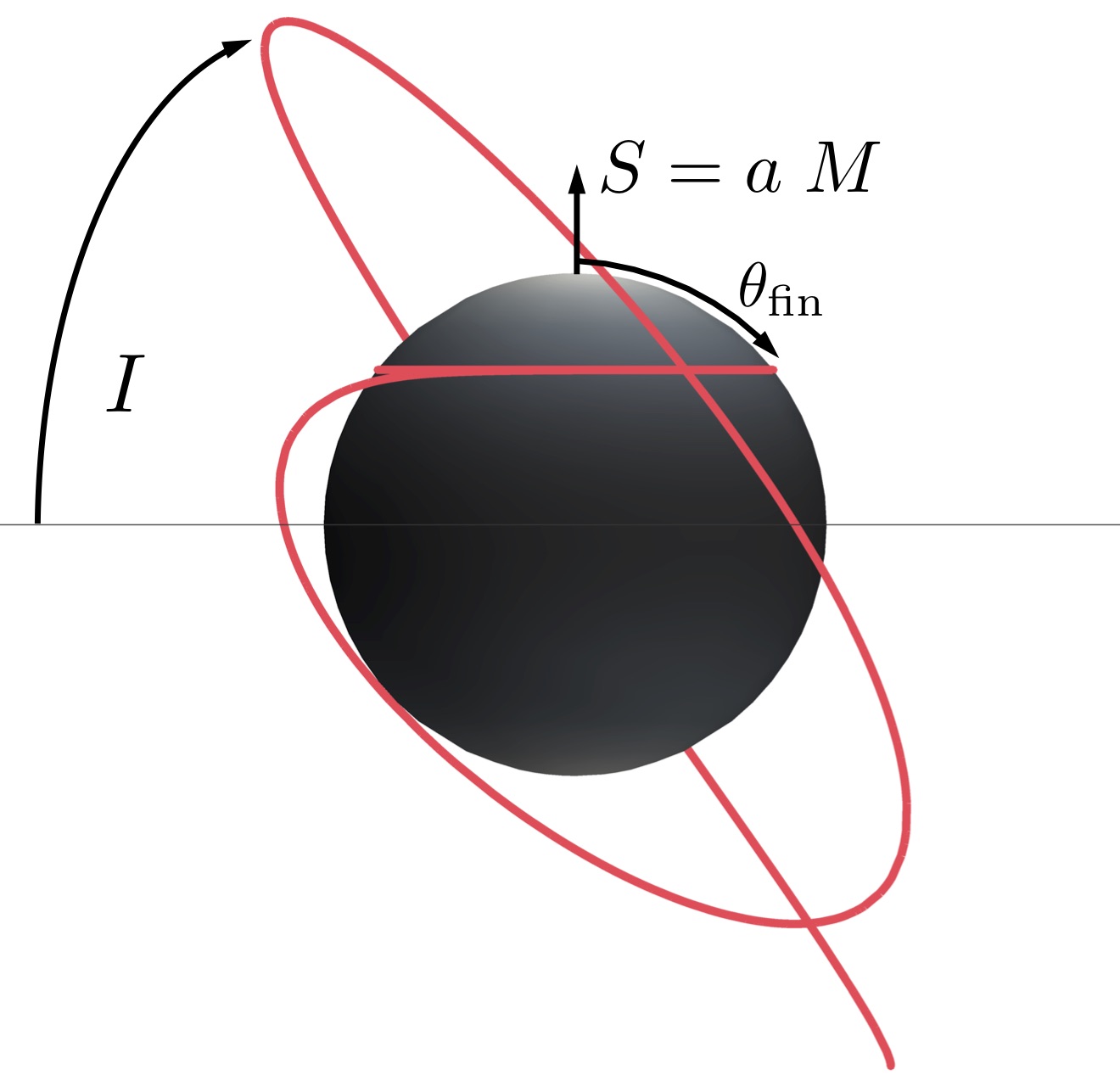}\\
\includegraphics[width = 0.2527\textwidth]{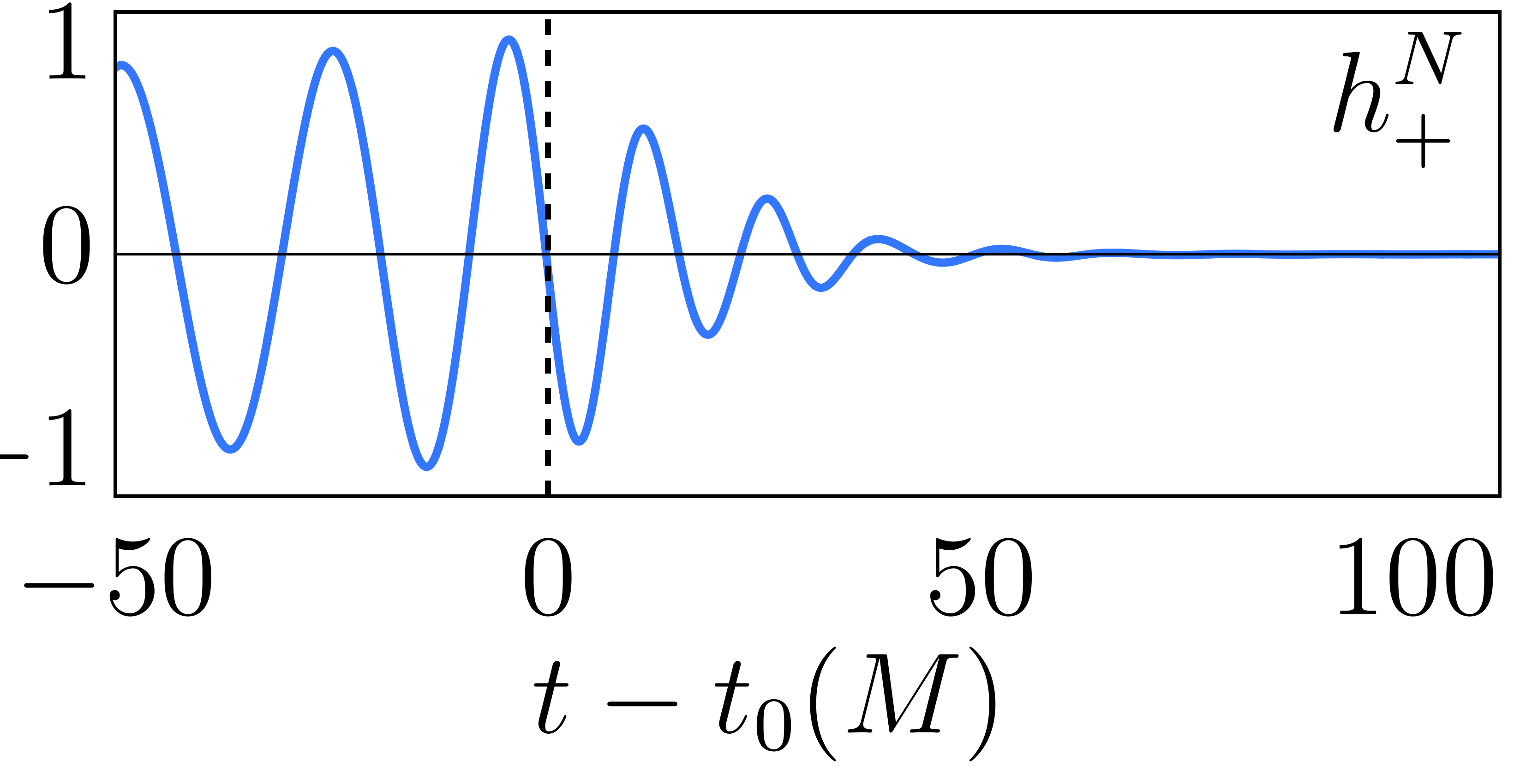}\qquad\qquad
\includegraphics[width = 0.2527\textwidth]{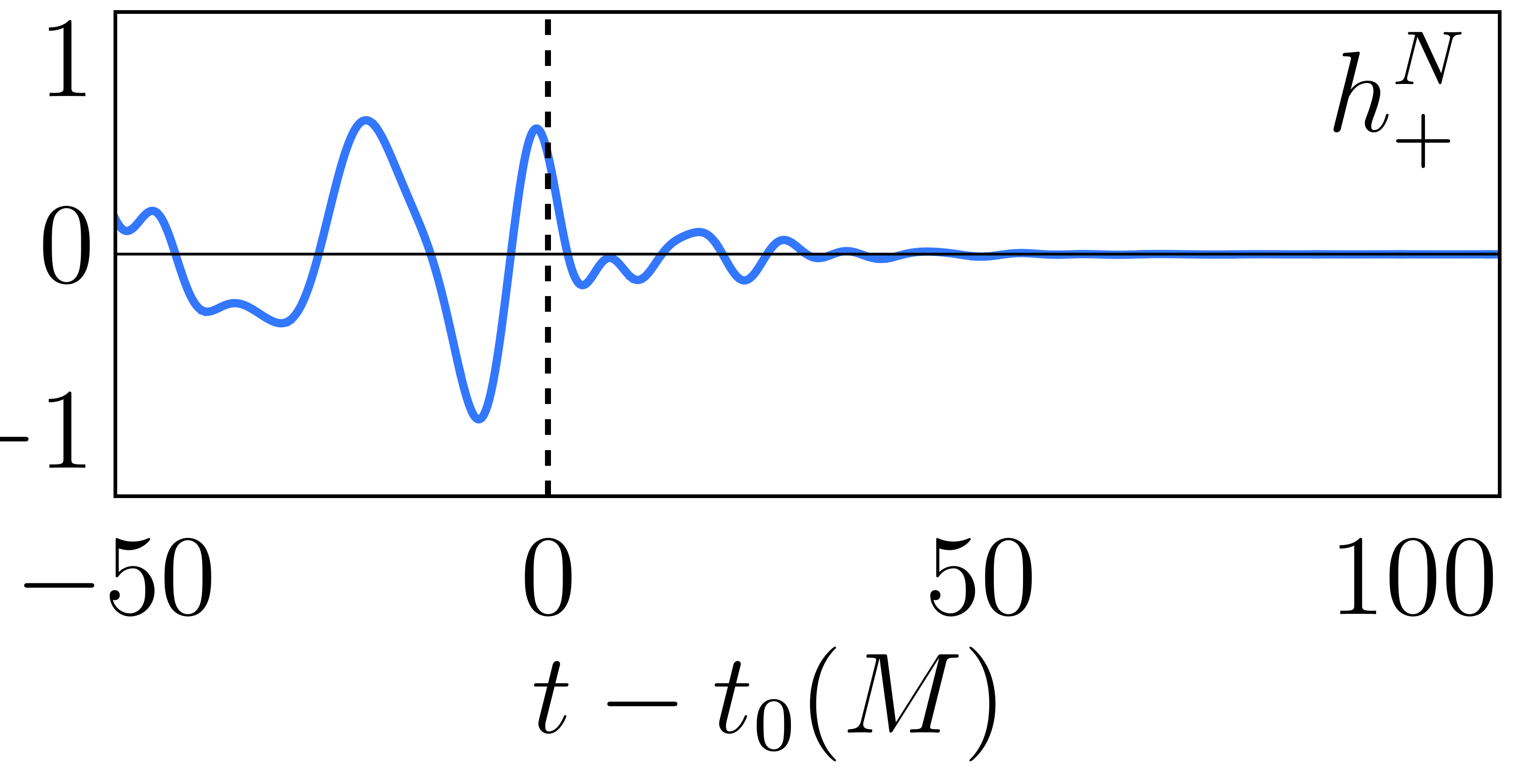}\\
\includegraphics[width = 0.2527\textwidth]{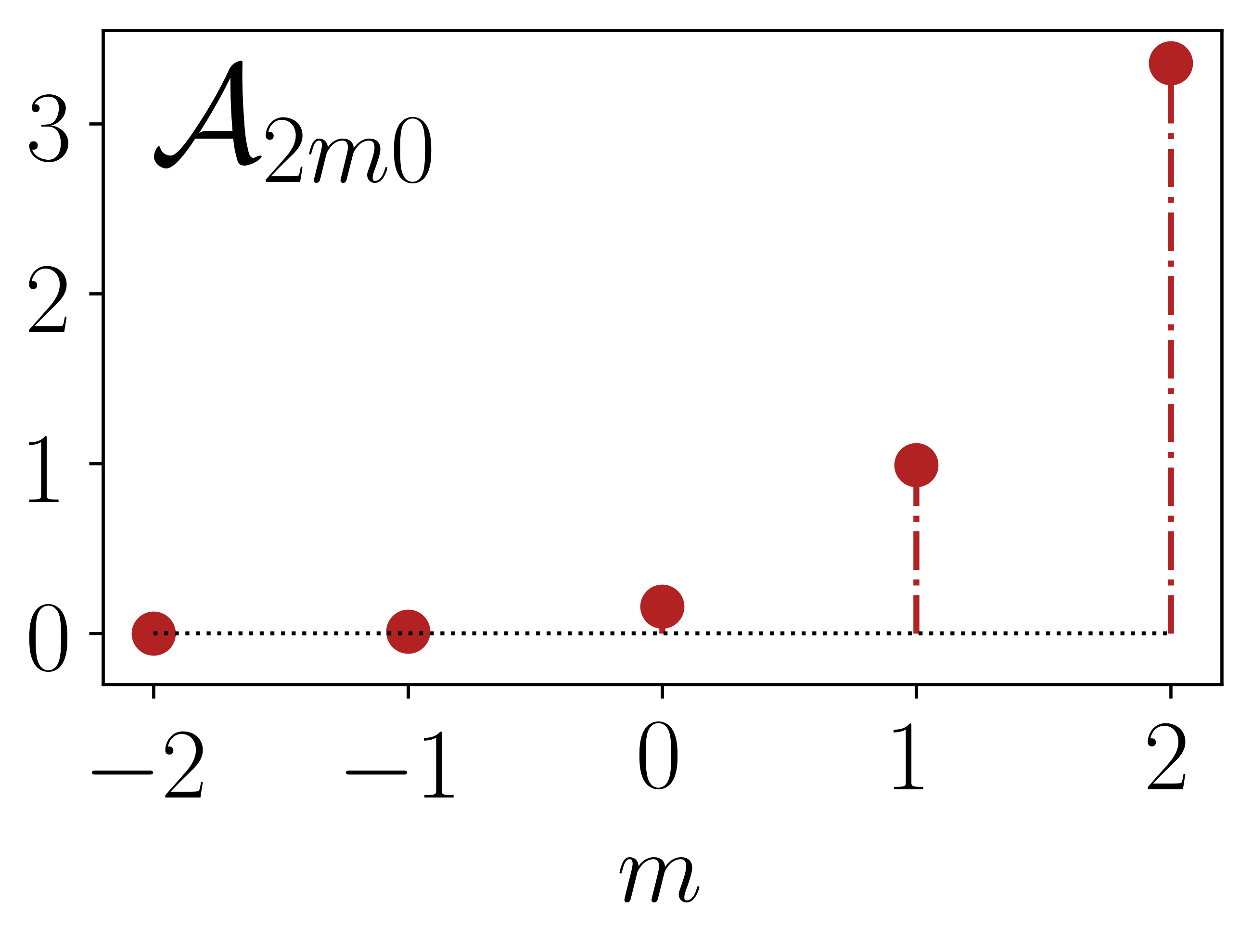}\qquad\qquad
\includegraphics[width = 0.2527\textwidth]{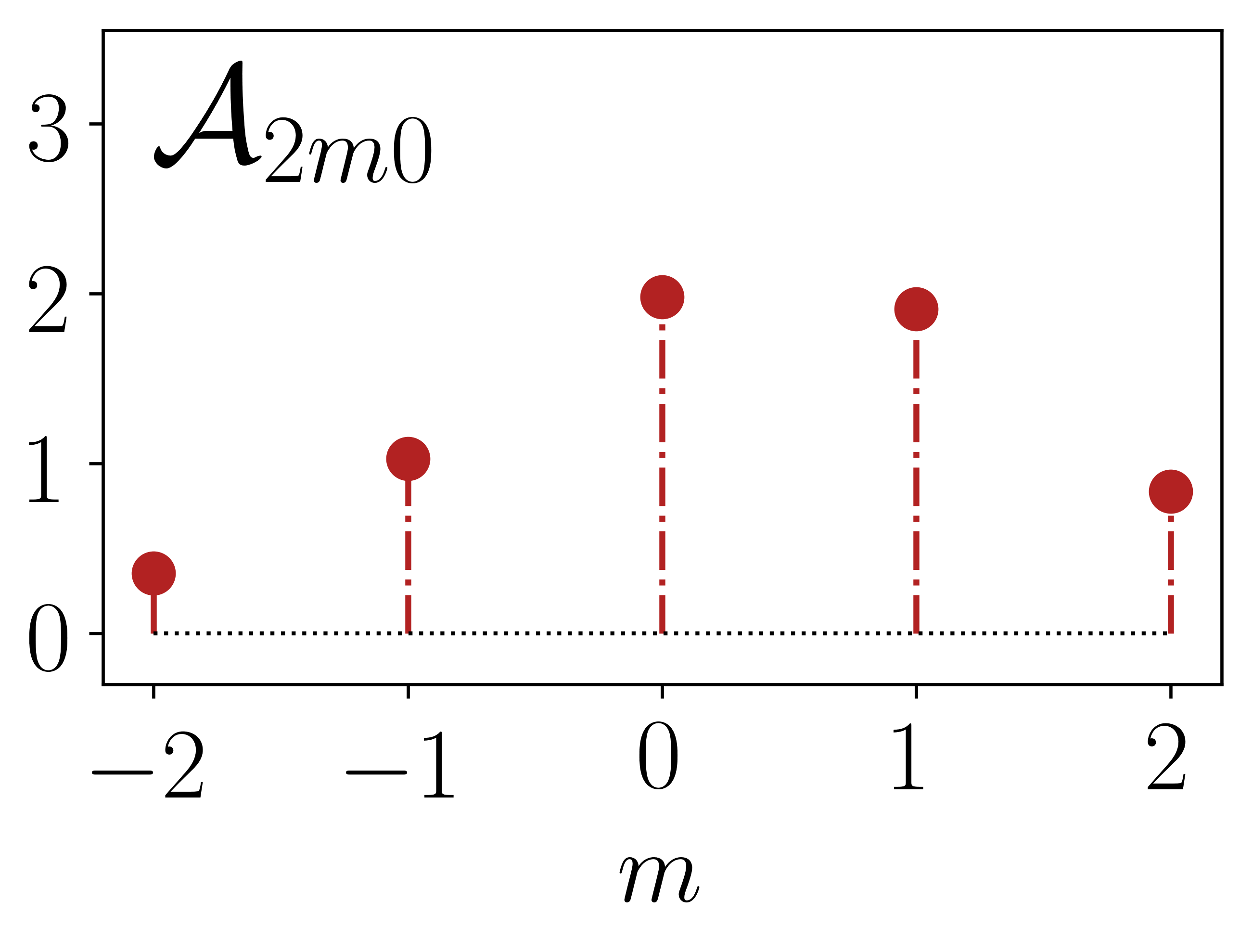}
\vskip -3mm
\caption{Two inspirals and plunges into a Kerr black hole (with spin $a = 0.5M$), the final waveform cycles that result, and some of the waveforms' mode content. Panels on the left describe equatorial inspiral and plunge; those on the right are for an orbit inclined at $I = 60^\circ$ to the equatorial plane.  Top panels show segments of the orbits' trajectories.  On the left, the small body is confined to $\theta = 90^\circ$, so not much detail is seen in the red worldline trace.  On the right, the small body's motion takes it through a range $30^\circ \le \theta \le 150^\circ$, plunging into the horizon at $\theta_{\rm fin} = 52^\circ$.  The middle panels show the final wave cycles ($+$ polarization in the equatorial plane) generated by these trajectories.  Notice the highly modulated waves resulting from the inclined orbit's more complicated motion.  Bottom shows the amplitudes of the $\ell = 2$ quasi-normal modes present in these waveforms [see Eq.\ (\ref{eq:spheroidaldecomp}) for the definition of ${\cal A}_{\ell m0}$].  The equatorial case is dominated by $m = 2$ and $m = 1$ modes; the inclined case has contributions from all allowed $m$.}
\label{fig:excite_geom}
\end{figure*}

Ringdown waves do not just describe the merged remnant, but also the mechanism that rings this black hole.  This information is encoded in the amplitudes of the different modes that are excited.  To our knowledge, the importance of this point for binary black holes was first argued in Refs.\ \cite{kamaretsos1,kamaretsos2}, in which the authors examined how information about a binary's mass ratio and its spins are encoded in the amplitudes of ringdown modes for the comparable mass case.  In our work, we consider large mass-ratio systems with orbits which can be highly misaligned from the larger black hole's spin.  Figure \ref{fig:excite_geom} illustrates what we find, contrasting ringdown generated by an aligned merger with a misaligned one.  The black hole rings in a manner akin to a bell that rings differently when struck differently.

The purpose of this {\it Letter} is to report recent findings in our work to develop a map between orbit geometry and ringdown mode excitation for large mass-ratio binaries {\cite{ah2019,lkah2019}}, and to argue that such a map could enhance what we learn about binary black holes from the ringdown (although more work will be needed to assess how our findings carry over to systems of less extreme mass ratio).  Information learned from these modes can complement what we learn about a system from the lower-frequency ``inspiral'' waveform.  The inspiral is well known to encode information about the orbit's geometry via spin-orbit precession effects, which modulate the inspiral's amplitude and frequency, slowly accumulating over many orbits.  Our work shows that a relatively high-frequency piece of the waveform also encodes this information.  To the extent that this finding carries over to less extreme mass ratios, this behavior for the ringdown waves may be particularly useful for systems at which most of the inspiral is at frequencies where detector sensitivity is relatively poor.  For current detectors, this is the case when the binary's total mass $M \gtrsim 50\,M_\odot$, a mass range that dominates the data so far {\cite{gwtc1}}.  The prospects are even richer for future ground- and space-based detectors currently under development {\cite{einsteintel,thirdgensens1,dwyeretal,thirdgensens2,rana,yulowfreq,lisa}}.

\smallskip
\noindent {\it Exciting and extracting black hole modes.}  We focus on the large mass-ratio limit, considering binaries in which one member (a Kerr black hole) is far more massive than the other.  This limits the immediate astrophysical applicability of our results, but allows us to study this problem using black hole perturbation theory {\cite{teuk}}.  This framework casts this problem as a particularly clean limit of the general relativistic two-body body, allowing us to compute many important quantities with very high precision.  In our concluding discussion, we discuss what our results imply for the general two-body problem and its potential importance for gravitational-wave astronomy.

A large mass-ratio binary's GWs can be modeled in two steps: first compute the worldline followed by a small body on an evolving orbit about a black hole, then use that worldline to compute the GWs the small body's motion produces.  Our model for the worldline, described in detail in {\cite{ah2019}}, generalizes the procedure developed by Ori and Thorne {\cite{ot00}}.  The small body is taken to have zero spin, and to move along on an orbit that is initially circular, but inclined with respect to the black hole's equatorial plane.  Its motion is broken into three epochs: an early {\it inspiral}, in which it adiabatically evolves through a sequence of circular orbits, driven by the backreaction of GW emission; a {\it transition}, when its motion is no longer adiabatic as it approaches the separatrix between stable and unstable orbits; and a final {\it plunge}, in which the small body follows an infalling trajectory through the black hole's event horizon.  The reader is referred to Ref.\ {\cite{ah2019}} for details of the assumptions and approximations which go into this model.

Once the worldline ${\bf z}(t)$ (the small body's position as a function of time used by distant observers) is known, we build the source $T$ of the Teukolsky equation {\cite{teuk}}:
\begin{eqnarray}
\label{eq:teuk}
&&
-\left[\frac{(r^2 + a^2)^2 }{\Delta}-a^2\sin^2\theta\right]
        \partial^2_{t}\Psi
-\frac{4 M a r}{\Delta}
        \partial_{t}\partial_{\phi}\Psi \nonumber \\
&&+4\left[r-\frac{M(r^2-a^2)}{\Delta}+ia\cos\theta\right]
        \partial_t\Psi\nonumber\\  
&&
+\,\Delta^{2}\partial_r\left(\Delta^{-1}\partial_r\Psi\right)
+\frac{1}{\sin\theta}\partial_\theta
\left(\sin\theta\,\partial_\theta\Psi\right)+\nonumber\\
&& \left[\frac{1}{\sin^2\theta}-\frac{a^2}{\Delta}\right] 
\partial^2_{\phi}\Psi - 4 \left[\frac{a (r-M)}{\Delta} 
+ \frac{i \cos\theta}{\sin^2\theta}\right] \partial_\phi\Psi  \nonumber\\
&&- \left(4 \cot^2\theta + 2 \right) \Psi = -4\pi\left(r^2+a^2\cos^2\theta\right)T\boldsymbol{(}{\bf z}(t)\boldsymbol{)}\;.
\end{eqnarray}
Here, $a \equiv |{\bf S}|/M$, the large black hole's spin divided by its mass, and $\Delta = r^2 - 2Mr + a^2$ (in units with $G = 1 = c$).  We set spin-weight to $-2$, for which $\Psi$ is simply related to the binary's GWs as measured by a distant observer.

An important property of the source $T$ in Eq.\ (\ref{eq:teuk}) is that it redshifts to zero as the small body approaches the event horizon.  The solutions to (\ref{eq:teuk}) for $T = 0$ are the quasi-normal modes describing black hole ringing.  This redshifting ensures that this model's last GW cycles are the hole's ringing modes, generated phase coherently with the preceding inspiral and plunge waves.  The waveforms shown in the middle row of Fig.\ {\ref{fig:excite_geom}} were produced by computing worldlines for inspiral and plunge (following the procedure described in Ref.\ {\cite{ah2019}}), building the source term $T$ for these cases, and then solving Eq.\ (\ref{eq:teuk}) using the code and techniques described in Refs.\ {\cite{td1,td2,td4}}.

Once inspiral and plunge GWs have been computed, we characterize their quasi-normal mode content using the procedure described in Ref.\ {\cite{lkah2019}}.  In brief, we use the fact that black hole quasi-normal mode GWs take the functional form {\cite{qnmlisa}}
\begin{eqnarray}
\!\!\!\!h(t) &=& \sum_{\ell mn} \bigg[\mathcal{A}_{\ell mn}e^{-i[\sigma_{\ell mn}(t-t_0) - \varphi_{\ell mn}]} S^{a\sigma_{\ell mn}}_{\ell mn}(\theta,\phi)
\nonumber\\ 
&+& \mathcal{A}'_{\ell mn} e^{i[\sigma^*_{\ell mn}(t-t_0)+\varphi'_{\ell mn}]}S^{a\sigma_{\ell mn}}_{\ell mn}(\pi-\theta,\phi)^*\bigg]\;.
\label{eq:spheroidaldecomp}
\end{eqnarray}
The time $t_0$ labels when the waveform becomes ringdown dominated, and $^*$ denotes complex conjugate.  Each mode's frequency $\omega_{\ell mn}$, damping time $\tau_{\ell mn}$ (with $\sigma_{\ell mn} = \omega_{\ell mn} - i/\tau_{\ell mn}$), and angular dependence $S^{a\sigma_{\ell mn}}_{\ell mn}(\theta,\phi)$ are well understood given black hole spin $a$, the angular indices $(\ell, m)$, and overtone number $n$ {\cite{qnmreview}}.  The amplitude ${\cal A}_{\ell mn}$ and phase $\varphi_{\ell mn}$ are extracted from the perturbation theory waveform using the algorithm described in Sec.\ III\,C of Ref.\ {\cite{lkah2019}}.  (The Kerr spacetime's symmetries enforce a simple relation between ${\cal A}_{\ell mn}$ and ${\cal A}'_{\ell mn}$ and between $\varphi_{\ell mn}$ and $\varphi'_{\ell mn}$.)  These amplitudes and phases fully characterize how the mode is excited by the late-time motion of a large mass-ratio binary.  By computing $({\cal A}_{\ell mn},\varphi_{\ell mn})$ for many modes and for different binary configurations, we build a map between binary properties and the excitation of each mode.  The bottom panels of Fig.\ {\ref{fig:excite_geom}} show ${\cal A}_{2m0}$ for the waveforms presented there.

\smallskip
\noindent {\it Our results.} Our main finding is that, at least in the large mass-ratio limit, there is a clean map between binary geometry and the ringing modes that dominate its final GW cycles, and that it is cleanly described by a small number of parameters which characterize the binary's final plunge geometry.  Given a black hole of mass $M$ and spin $a$, we find that the relative excitation of each $(\ell, m)$ mode (focusing on the ``fundamental'' modes, with $n = 0$) depends on two angles: the inclination $I$ that the small body's orbit makes with the hole's equatorial plane, and the polar angle $\theta_{\rm fin}$ at which the plunging small body crosses the event horizon.  Further discussion of these angles can be found in Ref. [8].  Orbits with $0^\circ \le I < 90^\circ$ are ``prograde,'' with axial angular momentum aligned to the black hole's spin; those with $90^\circ < I \le 180^\circ$ are ``retrograde,'' with antialigned axial angular momentum.  Equatorial orbits have $I = 0^\circ$ or $I = 180^\circ$.  Note that $I$ is nearly constant during an inspiral, typically increasing by a tenth of a degree or less for the cases we study.  Each mode's absolute excitation depends in addition on the small body's mass $\mu$ and the distance $D$ to the binary: $\mathcal{A}_{\ell mn} \propto \mu/D$.  In the results we present, we set this factor to 1; one can then scale the amplitudes by multiplying by $\mu/D$.

Note that $\theta_{\rm fin}$ is essentially an accidental phase parameter: given $I$, $\theta_{\rm fin}$ can be anywhere in $90^\circ - I \le \theta_{\rm fin} \le 90^\circ + I$ (for $I \le 90^\circ$) or $I - 90^\circ \le \theta_{\rm fin} \le 270^\circ - I$ (for $I > 90^\circ$).  Where the small body lands in this range depends on the relative phase of the binary's axial and polar motions.  The angle $\theta_{\rm fin}$ is thus a residue of the binary's initial conditions.  When we began this study, we first used an angle that described the initial polar phase of the worldline as our additional angle, and found that our results depended very strongly on parameters like the starting radius of the worldline.  It was only when we parameterized in terms of the worldline's {\it final} conditions that the ringdown spectrum showed predictable functional dependence on these angles.  In our conclusions, we argue that a similar dependence on accidental parameters must enter in the general case.

\begin{figure}[h]
\includegraphics[width = 0.27\textwidth]{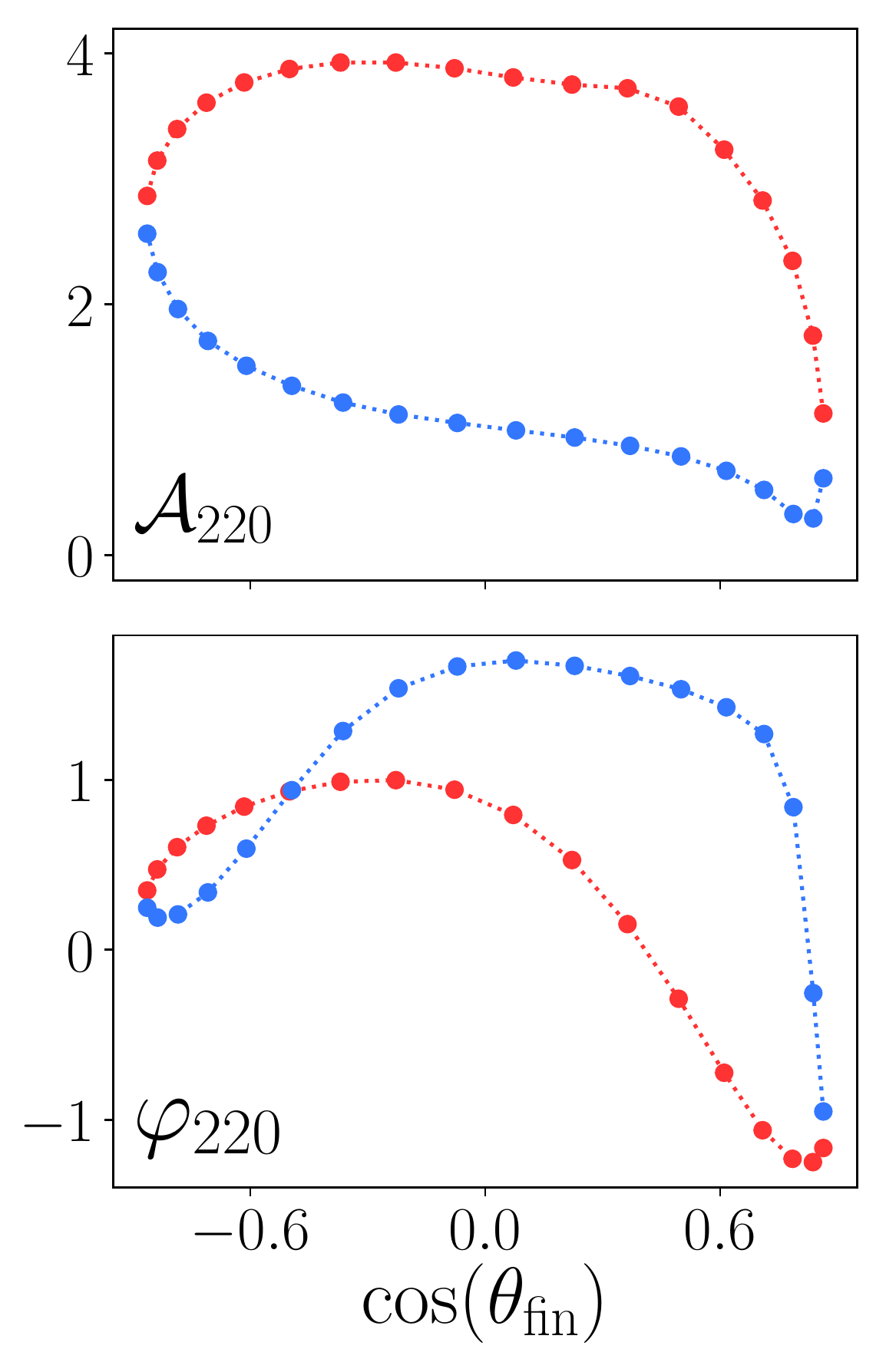}
\caption{An example of how mode excitation varies as a function of $\theta_{\rm fin}$, the polar angle at which the small body enters the event horizon.  Both panels are for $\ell = m = 2$ fundamental ($n = 0$) mode excitation of a black hole with $a = 0.5M$, and for orbit inclination $I = 60^\circ$.  Top panel shows the mode's amplitude ${\cal A}_{220}$ as a function of $\theta_{\rm fin}$; bottom shows its phase $\varphi_{220}$ in radians.  In both plots, the red dots are cases for which the plunging body has $\dot\theta_{\rm fin} > 0$, blue dots have $\dot\theta_{\rm fin} < 0$.}
\label{fig:ampphase_example}
\end{figure}

\begin{figure}[h]
\includegraphics[width = 0.34\textwidth]{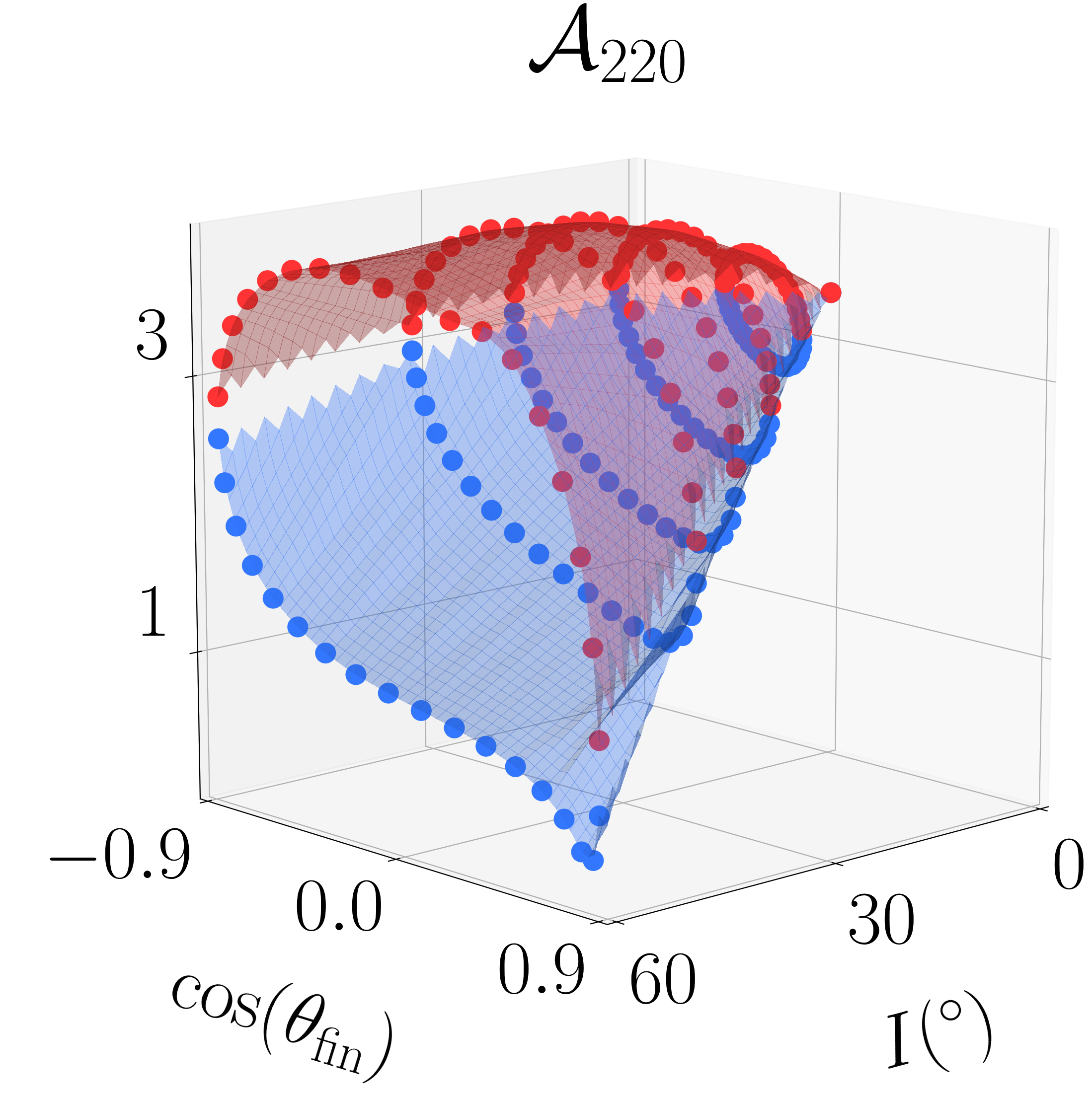}
\caption{Behavior of ${\cal A}_{220}$ for a black hole with $a = 0.5M$ as a function of orbital inclination $I$ and final polar angle $\theta_{\rm fin}$.  As in Fig.\ {\ref{fig:ampphase_example}}, red (blue) data is for $\dot\theta_{\rm fin} > 0$ ($\dot\theta_{\rm fin} < 0$).  This surface is quite smooth and well behaved.  Each orbit geometry cleanly picks out a particular mode amplitude.}
\label{fig:surface}
\end{figure}

Figure {\ref{fig:ampphase_example}} illustrates how a mode's amplitude and phase depends on $\theta_{\rm fin}$.  We show ${\cal A}_{220}$ and $\varphi_{220}$ for inspiral and plunge for a trajectory that is inclined $I = 60^\circ$ to the equatorial plane of an $a = 0.5M$ black hole.  The two tracks we show correspond to whether the small body's polar angle is decreasing when it crosses the horizon ($\dot\theta_{\rm fin} < 0$) or increasing ($\dot\theta_{\rm fin} > 0$).  These figures can be considered a slice of a surface traced out in $(I,\theta_{\rm fin})$; Fig.\ {\ref{fig:surface}} shows the surface for ${\cal A}_{220}$ over the range $0 \le I \le 60^\circ$ for all allowed $\theta_{\rm fin}$.  This surface is smooth and well behaved, cleanly mapping how mode characteristics depend on the orbit geometry.  Such figures become more baroque for other modes and other black hole spins.  Additional examples are shown in Sec.\ V\,A and Appendix A of Ref.\ {\cite{lkah2019}}.  Future work will explore the mode excitations and provide fits for the behavior that we find.

\noindent {\it Conclusions and discussion.} Our results show that, at least in the large mass-ratio limit, a binary's final ringdown waves describe not only the binary's merged remnant, but also its orbital geometry.  If this result carries over to less extreme mass ratios, it suggests that measuring multiple modes could make it possible to learn about spin-orbit misalignment, much as Refs.\ \cite{kamaretsos1,kamaretsos2} demonstrated that ringdown preserves memory of binary mass ratio.  Information about orbit geometry is much sought-after, since spin-orbit misalignment strongly depends upon the binary's formation mechanism (see, for example, Ref.\ {\cite{mandelfarmer2018}} for overview discussion, {\cite{ligoastro2016}} and references therein for discussion of these observables and source astrophysics, and {\cite{gerosaetal2018}} for recent discussion).  The imprint of misalignment-induced precession on the early inspiral may be hard to measure in many systems; ringdown may provide valuable complementary data in such cases.  In cases where both inspiral and ringdown probe the orbit geometry, one may be able to formulate a consistency test which will serve as a systematic check on the global goodness of fit for the GR-based models used to enable these measurements.

Additional work is needed to ascertain whether these results can be exploited for gravitational-wave astronomy.  On the measurement side, it will be important to understand how well multiple modes can be measured by detectors currently or soon to be operating.  Similar work has been done to understand how well mode frequencies and damping times can be measured {\cite{bhspectro1}}.  It will be important to determine how well the amplitudes of these modes can be measured, and how many modes must be measured and with what signal-to-noise ratio in order to learn about the binary geometry.  It will also be important to assess how strongly correlated modes are with each other and with other aspects of the signal.  Each mode is a damped sinusoid, so the likelihood that modes may be confused with one another is surely high.

Perhaps even more important will be to ascertain how well this result carries over to a wider range of binary black holes.  Although the large mass-ratio limit is a useful tool for exciting ringdown and characterizing how these modes depend on excitation geometry, it does not describe the physics of binaries currently being measured by gravitational-wave detectors.  As an example of physics missed by a large mass-ratio analysis, note that the system's final angular momentum is dominated by the black hole spin in our study.  Near mass ratio unity, the final angular momentum is dominated by the binary's orbit at plunge.  This will affect how to define the basis states which characterize quasi-normal modes.

For near unity mass ratios, the remnant's spin will be largely due to the orbital angular momentum, and so will be nearly parallel to the orbit.  Information about binary geometry will be washed away by plunge and merger dynamics; the resulting ringdown will likely be much like those we find for equatorial orbits.  There must be some mass ratio at which ringdown behavior similar to what we find first appears.  Numerical relativity will be needed to determine this mass ratio, and to explore the mapping between ringdown and the binary's geometry.  One lesson of our analysis is that one should expect the results to depend on an accidental parameter, akin to our plunge angle $\theta_{\rm fin}$ (although the most useful parameterization is likely to be more complicated away from the large mass-ratio limit).  A broad survey of this behavior may make it possible to construct phenomenological ringdown models describing the contributions of multiple modes to misaligned coalescences.

\section*{Acknowledgments}

This work was supported at MIT by NSF Grants PHY-1403261 and PHY-1707549; H.\ L.\ was also supported by an MIT Dean of Science Graduate Fellowship.  G.\ K.\ acknowledges research support from NSF Grant PHY-1701284 and ONR/DURIP Grant No.\ N00014181255.  We thank one of the referees for suggesting the inspiral/ringdown consistency test.

\bibliographystyle{apsrev4-1}
\bibliography{references}

\end{document}